\documentclass{elsart}

\usepackage{amssymb}
\usepackage{graphicx}

\journal{Optical Communication}
\begin{document}

\begin{frontmatter}

\title{A discrete fractional random transform}
\author{Zhengjun Liu, Haifa Zhao,}
\author{\ Shutian Liu\corauthref{cor1}}

\ead{stliu@hit.edu.cn} \corauth[cor1]{Corresponding author}
\address{Harbin Institute of Technology, Department of Physics, Harbin 150001 P. R. CHINA}

\begin{abstract}

We propose a discrete fractional random transform based on a
generalization of the discrete fractional Fourier transform with
an intrinsic randomness. Such discrete fractional random transform
inheres excellent mathematical properties of the fractional
Fourier transform along with some fantastic features of its own.
As a primary application, the discrete fractional random transform
has been used for image encryption and decryption.

\end{abstract}

\end{frontmatter}

\section{Introduction}

It is well known that the mathematical transforms from time (or
space) to frequency domain or joint time-frequency domain, such as
Fourier transform, Winger distribution function, wavelet transform
and more recent fractional Fourier transform, \textit{etc.} have
long been powerful mathematical tools in physics and information
processing. For instance, Fourier transform has been the basic tool
for signal representation, analysis and processing, image processing
and pattern recognition. In physics the Fourier transform describes
well the Fraunhofer (far field) diffraction of light and thus has
been the fundamental of information optics \cite{Goodman}. More
recently, in the research of quantum information, Fourier transform
algorithm has been adopted as an effective and fundamental algorithm
in quantum computer \cite{Nielsen}. Wavelet transform is a kind of
windowed Fourier transform (Gabor transform), however with variable
size of the windows \cite{wavelet}. Therefore wavelet transform has
been an extremely powerful tool in signal representation in
time-frequency joint domain with multi-resolution capability, and
has been extensively used in image compression, segmentation, fusion
and optical pattern recognitions.

The significance of mathematical transforms manifests itself
further when the fractional Fourier transform was re-invented in
1980's \cite{namias,mcbridge} and became active from 1990's after
its physical interpretations were found in optics
\cite{ozaktas1,lohmann,ozaktas2}. Actually at the beginning,
Namias tried to solve Schr{\"o}dinger equations in quantum
mechanics using fractional Fourier transform as a tool with little
notice in community. However, the fractional Fourier transform has
 found itself in real physical processes of light propagation
in a graded index (GRIN) fiber, which is equivalent to a near field
diffraction of light, Fresnel diffraction, with a quadratic factor.
Thus the fractional Fourier transform can be easily realized in a
bulk optical setup consisting of lenses. The fractional Fourier
transform provides various of new mathematical operations which are
useful in the field of optical information processing. And because
fractional Fourier transform also is a kind of time-frequency joint
representation of a signal \cite{almeida}, it has found extensive
applications in signal and image processing \cite{ozaktas3}.

The discrete forms of mathematical transforms have been extremely
useful in applications, especially in signal processing and image
manipulations. In fact, discrete transforms can approximate their
continuous versions with high precision, meanwhile with high
computation speed and lower complexities. Needless to say,
Discrete Fourier transform (or FFT) and discrete wavelet transform
have been widely used in different kinds of applications.
Recently, discrete fractional Fourier transform (DFrFT) and the
relevant discrete fractional cosine transform (DFrCT) have been
proposed \cite{dfrft,dfrct}. We have used this fast algorithm of
fractional Fourier transform in the numerical simulations of image
encryption and optical security \cite{encryp1,encryp2}.

As we have demonstrated that the extension of fractional Fourier
transform have many different kinds of definitions according to
how we fractionalize the Fourier transform \cite{genfrft}, the
DFrFT may also have different kinds of versions. In our researches
of optical image encryption, we ask naturally the question, is
there any possibility that the DFrFT be random? We have been
motivated in searching such a random transform because then the
image encryption process could be simplified by a single step of
transform. Recently we found that, from the generalization of
DFrFT, we can construct a discrete fractional random transform
(DFRNT) with an inherent randomness. We demonstrate that such
DFRNT has excellent mathematical properties as the fractional
Fourier transforms have. And moreover it has some fantastic
features of its own. We have also demonstrated that the DFRNT is a
very efficient tool in digital image encryption and decryption
with a very high speed of computation. The open questions left are
concerning with the physical analogies of DFRNT and its further
applications, which we are considering now.

We discuss the mathematical definition and properties of DFRNT and
provide numerical simulation results of the DFRNT's for
one-dimensional and two-dimensional signals in the following
sections in details.

\section{Mathematics of the discrete fractional random transform}

We begin our discussions from the definition of DFrFT proposed by
Pei {\it et al} \cite{dfrft}. A one-dimensional DFrFT can be
expressed as a matrix-vector multiplication
\begin{equation}
\mathbf{X}_{\alpha}(n)=\mathbf{F}^{\alpha}\mathbf{x}(n),
\end{equation}
where $\mathbf{x}(n)$ is the input vector which has $N$ elements,
$\mathbf{F}^{\alpha}$ is the kernel transform matrix and $\alpha$ is
the fractional order. When $\alpha=1$, the DFrFT becomes the DFT as
$\mathbf{X}(n)=\mathbf{F}\mathbf{x}(n)$, with matrix $\mathbf{F}$
indicating the kernel matrix of DFT.

The transform matrix is defined as follows. Firstly, because the
fractional Fourier transform has the same eigenfunctions with the
Fourier transform, we can calculate the eigenvectors
$\{\mathbf{V}_j\}$ ($j=1,2,\dots,N$) of DFrFT with a real
transform matrix $\mathbf{S}$ of discrete Fourier transform which
is defined as follows \cite{dft}
\begin{equation}
\mathbf{S}=\left[
\begin{array}{ccccccc}
2 & 1 & 0 & 0 & \dots & 0 & 1 \\
1 & 2\cos \omega & 1 & 0 & \dots & 0 & 0 \\
0 & 1 & 2 \cos 2\omega & 1 & \dots & 0 & 0\\
\vdots & \vdots & \vdots & \ddots & \vdots & \vdots & \vdots\\
1 & 0 & 0& 0 & \dots & 1 & 2\cos (N-1)\omega \end{array}\right],
\end{equation}
where $\omega=2\pi/N$. The matrix $\mathbf{S}$ actually is not the
kernel transform matrix $\mathbf{F}$ of the discrete Fourier
transform (DFT). However, the matrix $\mathbf{S}$ commutes with
matrix $\mathbf{F}$, \textit{i.e.}
$\mathbf{S}\mathbf{F}=\mathbf{F}\mathbf{S}$. Thus the eigenvectors
of $\mathbf{S}$ are also the eigenvectors of $\mathbf{F}$, only they
correspond to different eigenvalues. Because the matrix $\mathbf{S}$
is symmetrical, the eigenvectors \{$\mathbf{V}_j$\} are all real and
orthonormal to each other. They form an orthonormal basis which is
equivalent to the Hermite-Gaussian polynomials in the case of
continuous Fourier transform and fractional Fourier transform.
Therefore, from the matrix $\mathbf{S}$ we obtain the $N \times N$
eigenvector basis matrix $\mathbf{V}$, which is formed by the $N$
column eigenvectors
$\{\mathbf{v}_1,\mathbf{v}_2,\dots,\mathbf{v}_N\}$ as
$\mathbf{V}=\left[\mathbf{v}_1\; \mathbf{v}_2\; \dots\; \mathbf{v}_N
\right]$. In the calculation of DFrFT, the DFT-shifted version of
eigenvectors \{$\mathbf{V}_j$\} are taken.

Next step, we determine the eigenvalues of DFrFT. We know that the
eigenvalues of the continuous fractional Fourier transform can be
written as
\begin{equation}
\lambda_k=\exp(-i\alpha k \pi/2),\;k=0,1,2,\dots, \infty.
\end{equation}
In DFrFT the eigenvalues are not changed, only we have a limited
numbers of eigenvalues being taken into account, say
$k=0,1,2,\dots,N$. Those eigenvalues again construct a matrix
$\mathbf{D}_{\alpha}$ as follows
\begin{equation}
\mathbf{D}_{\alpha}=\left\{\begin{array}{llll}\mathrm{diag}\left(1,e^{-i
\alpha \pi/2},\dots,e^{-i \alpha (N-1) \pi/2}\right)&
 \mbox{if $N$ is odd,}\\
\mathrm{diag}\left(1,e^{-i \alpha \pi/2}, \dots,e^{-i \alpha (N-2)
\pi/2}, e^{-i \alpha N \pi/2}\right) & \mbox{if $N$ is even},
\end{array}\right.
\end{equation}
where $\mathbf{D}_{\alpha}$ is an $N \times N$ diagonal matrix. It
must be noted that in the expression of $\mathbf{D}_{\alpha}$,
there is a jump in the last eigenvalue for the $N$ is an even
integer. Such assignments of eigenvalues of DFrFT are consistent
with the DFT's multiplicity rules \cite{mccellan}.

So now we have already had the eigenvector basis $\mathbf{V}$ and
the corresponding eigenvalues $\mathbf{D}_{\alpha}$. In the final
step, we can express a kernel transform of DFrFT by the
eigen-decomposition. The transform matrix of the DFrFT can then be
defined as
\begin{equation}
{\mathbf F}^{\alpha}=\mathbf{V}\mathbf{D}_{\alpha}\mathbf{V}^{T},
\end{equation}
where $\mathbf{V}^T$ indicates the transpose of the matrix
$\mathbf{V}$. Because the eigenvectors are orthonormal, we have
$\mathbf{V}\mathbf{V}^{T}=\mathbf{I}$ and $\mathbf{I}$ is the
identity matrix. And also
$\mathbf{D}_{-\alpha}=\mathbf{D}_{\alpha}^{*}$. Those relations
conform that the DFrFT have the same properties with the
continuous FrFT.

The above procedure gives a formal way to construct the DFrFT. The
matrix ${\mathbf S}$ is the most important figure in the both
DFrFT and DFT, because from it the eigenvectors of DFT and DFrFT
can be calculated. The matrix ${\mathbf S}$ is symmetrical and
therefore its eigenvalues are all real and the eigenvectors are
orthonormal. The transform kernel matrix of DFT and DFrFT can thus
be constructed by eigen-decomposition with different eigenvalues.

How does a discrete fractional Fourier transform becomes random? The
essence of the generalization from the DFrFT to DFRNT is to change
the matrix $\mathbf{S}$ to a random matrix. The DFRNT can be defined
by a symmetric random matrix $\mathbf{Q}$. The matrix $\mathbf{Q}$
is generated by an $N \times N$ real random matrix $\mathbf{P}$ with
a relation of
\begin{equation}
\mathbf{Q}=(\mathbf{P}+\mathbf{P}^T)/2,
\end{equation}
where we have ${Q}_{lk}={Q}_{kl}$. Similar to the definition of
DFrFT, we can also generate $N$ real orthogonal eigenvectors
$\{\mathbf{v}'_{R1},\mathbf{v}'_{R2},\dots,\mathbf{v}'_{RN}\}$ of
matrix ${\mathbf Q}$. Those eigenvectors can be normalized by the
Schmidt standard normalization procedure. Then we have $N$
orthonormal eigenvectors
$\{\mathbf{v}_{R1},\mathbf{v}_{R2},\dots,\mathbf{v}_{RN}\}$. From
those column vectors a matrix
\begin{equation}
\mathbf{V}_R=\left[\mathbf{v}_{R1}\;\mathbf{v}_{R2}\;\dots\;\mathbf{v}_{RN}\right]
\end{equation}
can be achieved, where
$\mathbf{V}_{R}\mathbf{V}_{R}^{T}=\mathbf{I}$. We do not need to
DFT-shift the eigenvectors here, because these eigenvectors are
generated from a symmetrical random matrix.

The coefficient matrix, that corresponds to the eigenvalues of
DFRNT can be defined as
\begin{eqnarray}
\mathbf{D}_{R\alpha}&=&
\mathrm{diag}\left(1,\exp\left(-i\frac{2\pi
\alpha}{M}\right),\exp\left(-i\frac{4\pi
\alpha}{M}\right),\right.\nonumber \\
&&\quad \quad \quad \quad \quad \quad \quad \quad \quad \left.
\dots,\exp\left(-i\frac{2(N-1)\pi \alpha}{M}\right)\right),
\end{eqnarray}
which is used in the process of eigen-decomposition. One should
notice that the eigenvalues of DFRNT are not necessarily relevant
to those of DFrFT's or DFT's. However, we take a similar form so
that the DFRNT may have similar mathematical properties. In
Eq.~(8) there is no jump for odd and even integer $N$. We
introduce here an integer number $M$ in the coefficients. It
indicates the periodicity of DFRNT with respect to the fractional
order $\alpha$ whose significance will be shown bellow. The kernel
transform matrix of DFRNT can thus be expressed as
\begin{equation}
{\mathbf
R}^{\alpha}=\mathbf{V}_{R}\mathbf{D}_{R\alpha}\mathbf{V}^{T}_{R}.
\end{equation}
Therefore the DFRNT of a one-dimensional discrete signal is
written as
\begin{equation}
\mathbf{X}_{R(\alpha)}(n)=\mathbf{R}^{\alpha}\mathbf{x}(n)
\end{equation}
The expansion of DFRNT for two dimensional signal is
straightforward as $\mathbf{X}_{R(\alpha)
}=\mathbf{R}^{\alpha}\mathbf{x}\left(\mathbf{R}^{\alpha}\right)^{T}$.

The most important feature of DFRNT is that its transform kernel
is random, which results from the randomness of matrix
$\mathbf{Q}$, so that the result of transform is totally random.
The eigenvectors of DFRNT depends upon the random matrix
$\mathbf{Q}$ (or $\mathbf{P}$), therefore if we change the matrix
$\mathbf{P}$, then the results of DFRNT's is different.
Furthermore, the DFRNT inheres most of the excellent properties of
DFT (or DFrFT). It can be easily verified that the DFRNT has the
following mathematical properties.
\begin{itemize}
    \item \textbf{Linearity}, DFRNT is a linear transform,
    \textit{i.e.} $\mathbf{R}^{\alpha}(a\mathbf{x}+b\mathbf{y})=
    a\mathbf{R}^{\alpha}\mathbf{x}+b\mathbf{R}^{\alpha}\mathbf{y}$,
    where $a$ and $b$ are constants.
    \item \textbf{Unitarity}, DFRNT is a unitary transform, \textit{i.e.}
    $\mathbf{R}^{-\alpha}=\left(\mathbf{R}^{\alpha}\right)^{*}$ because
    $\mathbf{D}_{R(-\alpha)}=\mathbf{D}^{*}_{R\alpha}$. The
    inverse DFRNT exists and is defined as $\mathbf{R}^{-\alpha}$.
    \item \textbf{Additivity}, DFRNT obeys the additive
    role as the DFrFT (and FrFT) does for the fact that
    $\mathbf{R}^{\alpha}\mathbf{R}^{\beta}=
    \mathbf{R}^{\beta}\mathbf{R}^{\alpha}=\mathbf{R}^{\alpha+\beta}$.
    \item \textbf{Multiplicity}, DFRNT has a periodicity of $M$ in
    our definition, \textit{i.e.} $\mathbf{R}^{\alpha+M}=
    \mathbf{R}^{\alpha}$, where we can change this periodicity with
    changing the integer $M$.
    \item \textbf{Parseval}, DFRNT satisfy the Parseval energy conservation
    theorem, \textit{i.e.} $\sum_{k=0}^{N-1}\left|{X}_{\alpha
    (R)}(k)\right|^2=\sum_{m=0}^{N-1}\left|x(m)\right|^2$.
\end{itemize}

What we are interested most is randomness of DFRNT and the
information retrieval capability because of its multiplicity. As
the fractional order $\alpha=lM$, where $l$ is another integer,
the DFRNT output signal $\mathbf{X}_{R(\alpha)}$ return to its
original function $\mathbf{x}$. Otherwise, it will be totally
random when the order $\alpha\neq lM$ even a small aberration
occurs. While with the order $\alpha=lM/2$, \textit{i.e.} at the
half of its period, the output signal $\mathbf{X}_{R(\alpha)}$ is
real. Such a fascinating feature can be rigorously proved in
mathematics, because in this case $\mathbf{D}_{R\alpha}$ is real
(and therefore the kernel transform matrix $\mathbf{R}^{\alpha}$
is real). It may also be intuitively seen if one recall that the
Fourier transform has the property of
$\mathcal{F}^2\{f(x)\}=f(-x)$. The only difference is that the
amplitude of DFRNT, when $\alpha=lM/2$, is random but not
$\left|\mathbf{x}(-n)\right|$.

%
%

To illustrate the basic feature of DFRNT, and to make a comparison
with DFrFT, we present here the simulation results for
one-dimensional rectangular signal using DFrFT and DFRNT, in Fig.~1
to Fig.~3, respectively. The signal has a rectangular window with
period of $[40,60]$. The total number of points is $N=100$. The
numerical results of DFrFT is given in Fig.~1, with the fractional
order $\alpha=0.25,\;0.50,\;0.75$ and $1.00$, respectively. We can
see that the amplitudes of DFrFT's gradually change from the
original signal to its Fourier transform. However, the results will
be different for the cases of DFRNT. We calculate DFRNT with two
different formats of random numbers, one is normally distributed
random number (illustrated in Fig.~2) and the other is uniform
random number within $[0,1]$ (Fig.~3), as the matrix $\mathbf{P}$.
Here we set the periodicity of DFRNT as $M=1$. From the results
shown in Fig.~2 and Fig.~3, one can see that the amplitudes are all
randomly distributed when the fractional order $\alpha \neq 1$.
Whereas when $\alpha=1$, the output signal goes back to its
original, \textit{i.e.} the signal recovers.

Note that when $\alpha=0.50$, the phase values are $0$, $\pi$, or
$-\pi$ whatever random matrix is used. That means we always get
real transform results when the fractional order $\alpha$ is the
half of the periodicity $M$. The amplitudes for $\alpha=0.25$ and
$\alpha=0.75$ are the same, however their phases are conjugated.
That is because
$\mathbf{X}_{R(M-\alpha)}=\mathbf{X}_{R(-\alpha)}=\left(
\mathbf{X}_{R(\alpha)}\right)^{*}$. When $\alpha=1$, the
amplitudes of transformed signal totally retrieved, however, there
exit about $\pm \pi/2$ phase fluctuations at the components of
$x(k)=0$.

\section{Image encryption and decryption: a primary application of DFRNT}

The primary and perhaps the most important application of DFRNT is
cryptography and information security. Because the DFRNT itself is
random, the DFRNT of a two-dimensional signal can be directly used
for image encryption with any fractional order $\alpha\neq lM$.
The decryption process is simply an inverse DFRNT. The main
encryption key is the matrix $\mathbf{Q}$. We have known that
changing a random matrix $\mathbf{Q}$ indicates changing the
present DFRNT to another DFRNT. The results are unrelated even
they have the same fractional order $\alpha$. The DFRNT is quite
sensitive to the random matrix and also to the fractional order,
which have been demonstrated in our numerical simulations.

Numerical results of DFRNT's of an image are given in Fig.~4 and
Fig.~5 with normally and uniformly distributed random matrix
respectively. The test image is the photo of Lenna with the size
of $256 \times 256$ pixels, shown in Fig.~4(a) and Fig.~5(a). The
encryption process is simple, just an $\alpha$ order DFRNT with
$\alpha \neq lM$. The encryption results are shown in Fig.~4(b),
Fig.~4(c), Fig.~5(b) and Fig.~5(c), respectively, with
$\alpha=0.50$ and $\alpha=0.80$. In our simulations we let $M=1$.
The decryption is an inverse DFRNT with same random matrix
$\mathbf{Q}$ and the fractional order $-\alpha$. However, if we
perform the inverse DFRNT with another matrix $\mathbf{Q}$ which
is different from the matrix we use in the encryption process, we
fail to retrieve the image. The decryption results are shown in
Fig.~4(d) and Fig.~5(d). And also it is worth to mention that the
fractional order $\alpha$ can also be served as an extra
decryption key. The inverse DFNRT's with wrong fractional orders
can not recover the image as shown in Fig.~4(e) and Fig.~5(e),
with $\alpha=-0.502$ and $\alpha=-0.503$, respectively. As we have
analyzed from the computation of mean square error (MSE), that the
smallest security discrimination is about $|\Delta
\alpha|_{min}\approx 0.02$ for normally distributed random matrix
and $|\Delta \alpha|_{min}\approx 0.03$ for the case of uniformly
distributed random matrix. When $|\Delta \alpha|>|\Delta
\alpha|_{min}$, one can not visually recognize the image from the
noisy background. The correct decryption results are shown in
Fig.~4(f) and Fig.~5(f). The MSE for correct decryption process is
in the order of $10^{-13}$ for both cases.

The image encryption with DFRNT using the fractional order
$\alpha=lM/2$ is of great interest for real applications, because
then the encrypted image has real values. It may be very useful
for storing and transferring the secret data in a more convenient
way, for example, by a photo plate.

%

The security strength of the DFRNT encryption can be estimated as
$2^{N(N+1)/2}$ because the random matrix $\mathbf{Q}$ has $N(N+1)/2$
independent elements. As a matter of fact, if one try to search such
a random matrix coded with uniformly distributed random numbers, the
number of steps will be much larger than $2^{N(N+1)/2}$. Therefore,
such an image encryption method is considerably secure in theory.
This encryption algorithm can be easily implemented digitally.

Is there any applications of DFRNT in Physics? This question now
is left for an open problem. Nevertheless, DFRNT has a powerful
multiplicities with changing the formats of matrix $\mathbf{Q}$.
The different matrix $\mathbf{Q}$ may results in different
transform with different mathematical properties. Such a feature
may be helpful in filter designs in signal and image processing.
We can also establish an iterative mechanism that the matrix
$\mathbf{Q}$ can be modified accordingly. This mechanism may be
found applications in solving inverse problems in optics, such as
the phase retrieval.

\section{Conclusions}

We have proposed a new kind of discrete transform, a discrete
fractional random transform, based on a generalization of the
discrete fractional Fourier transform. The intrinsic randomness of
the discrete fractional random transform origins from a symmetrical
random matrix, from which the eigenvectors of the transform are
generated. The discrete fractional random transform inheres the
excellent mathematical properties as the fractional Fourier
transforms have. A new image encryption and decryption scheme is
proposed that uses the discrete fractional random transform. Thus
the processes of image encryption and decryption are very simple,
only consisting of an operation of discrete fractional random
transform and its inverse. Mathematical properties of the new
discrete transform have been given in details with numerical
demonstration.

\newpage

\section*{List of figure captions}

Figure 1. DFrFT of a one-dimensional rectangular window signal.
The fractional orders are $\alpha=0.25$, $\alpha=0.50$,
$\alpha=0.75$ and $\alpha=1.00$, respectively.

\vspace{1cm}\noindent Figure 2. DFRNT of a one-dimensional
rectangular window signal with normally distributed random
numbers. The fractional orders are $\alpha=0.25$, $\alpha=0.50$,
$\alpha=0.75$ and $\alpha=1.00$, respectively.

\vspace{1cm}\noindent Figure 3. DFRNT of a one-dimensional
rectangular window signal with uniformly distributed random
numbers. The fractional orders are $\alpha=0.25$, $\alpha=0.50$,
$\alpha=0.75$ and $\alpha=1.00$, respectively.

\vspace{1cm}\noindent Figure 4. Numerical results of DFRNT with
two-dimensional data using a normally distributed random matrix.
DFRNT serves as an image encryption and decryption algorithm here.
(a) The original image, (b) encrypted image with $\alpha=0.5$, (c)
encrypted image with $\alpha=0.8$, (d) decryption result for image
(b) with a different random matrix, (e) decryption result for
image (b) with $\alpha=-0.502$, and (f) the correct decryption of
the image.

\vspace{1cm}\noindent Figure 5. Numerical results of DFRNT with
two-dimensional data using a uniformly distributed random matrix.
(a) The original image, (b) encrypted image with $\alpha=0.5$, (c)
encrypted image with $\alpha=0.8$, (d) decryption result for image
(b) with a different random matrix, (e) decryption result for image
(b) with $\alpha=-0.503$, and (f) the correct decryption of the
image.

\newpage
\begin{figure}[h]
\includegraphics[width=17.50cm]{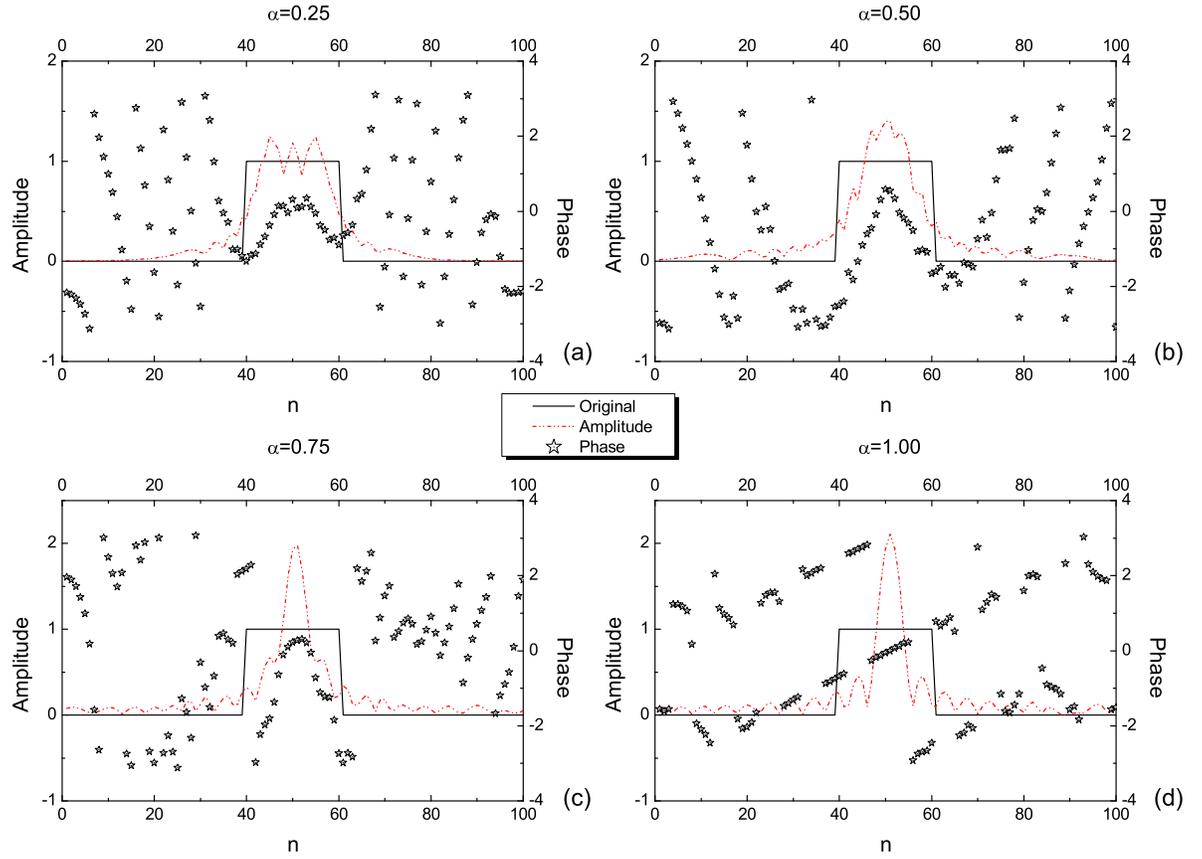}
  \caption{Numerical simulations of DFrFT of a one-dimensional rectangular
window signal. The fractional orders are $\alpha=0.25$,
$\alpha=0.50$, $\alpha=0.75$ and $\alpha=1.00$, respectively.}
  \end{figure}

\begin{figure}[h]
\includegraphics[width=17.50cm]{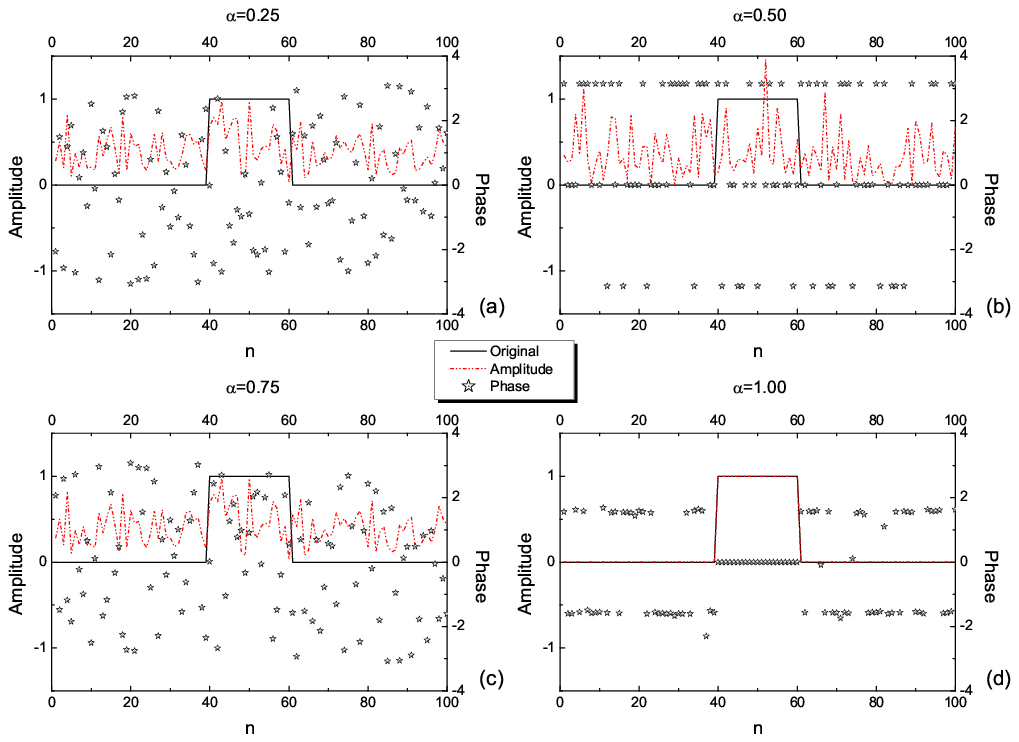}
  \caption{Numerical simulations of DFRNT of a one-dimensional rectangular
window signal with normally distributed random numbers. The
fractional orders are $\alpha=0.25$, $\alpha=0.50$, $\alpha=0.75$
and $\alpha=1.00$, respectively.}
  \end{figure}

\begin{figure}[h]
\includegraphics[width=17.50cm]{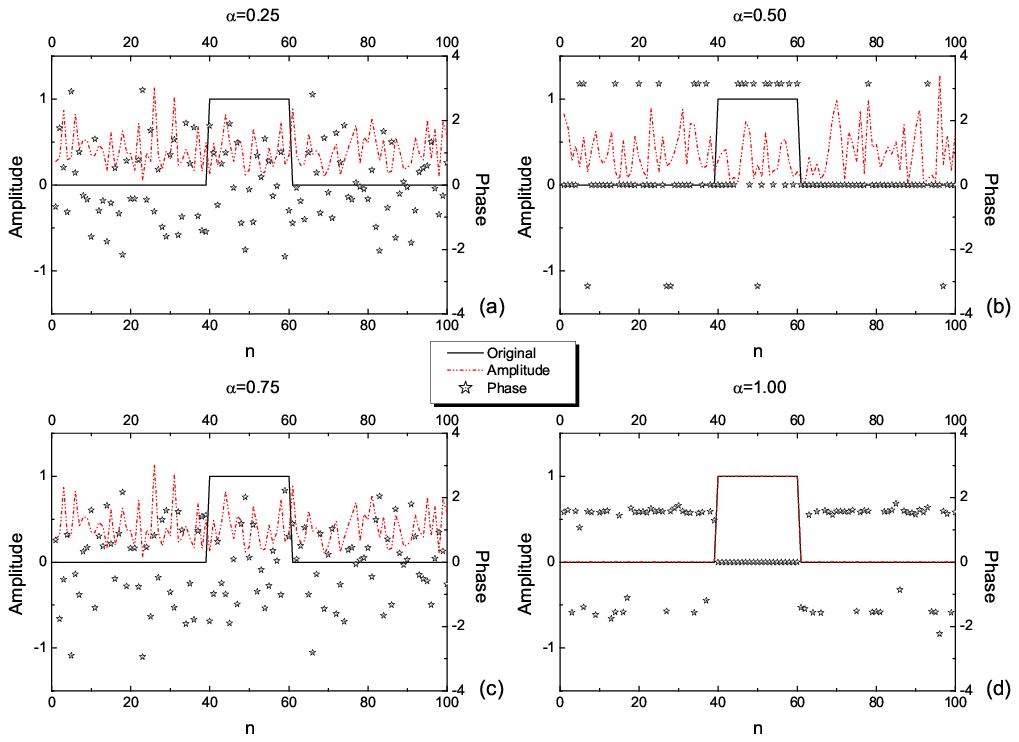}
  \caption{Numerical simulations of DFRNT of a one-dimensional rectangular
window signal with uniformly distributed random numbers. The
fractional orders are $\alpha=0.25$, $\alpha=0.50$, $\alpha=0.75$
and $\alpha=1.00$, respectively.}
  \end{figure}

\begin{figure}[h]
\includegraphics[width=16.00cm]{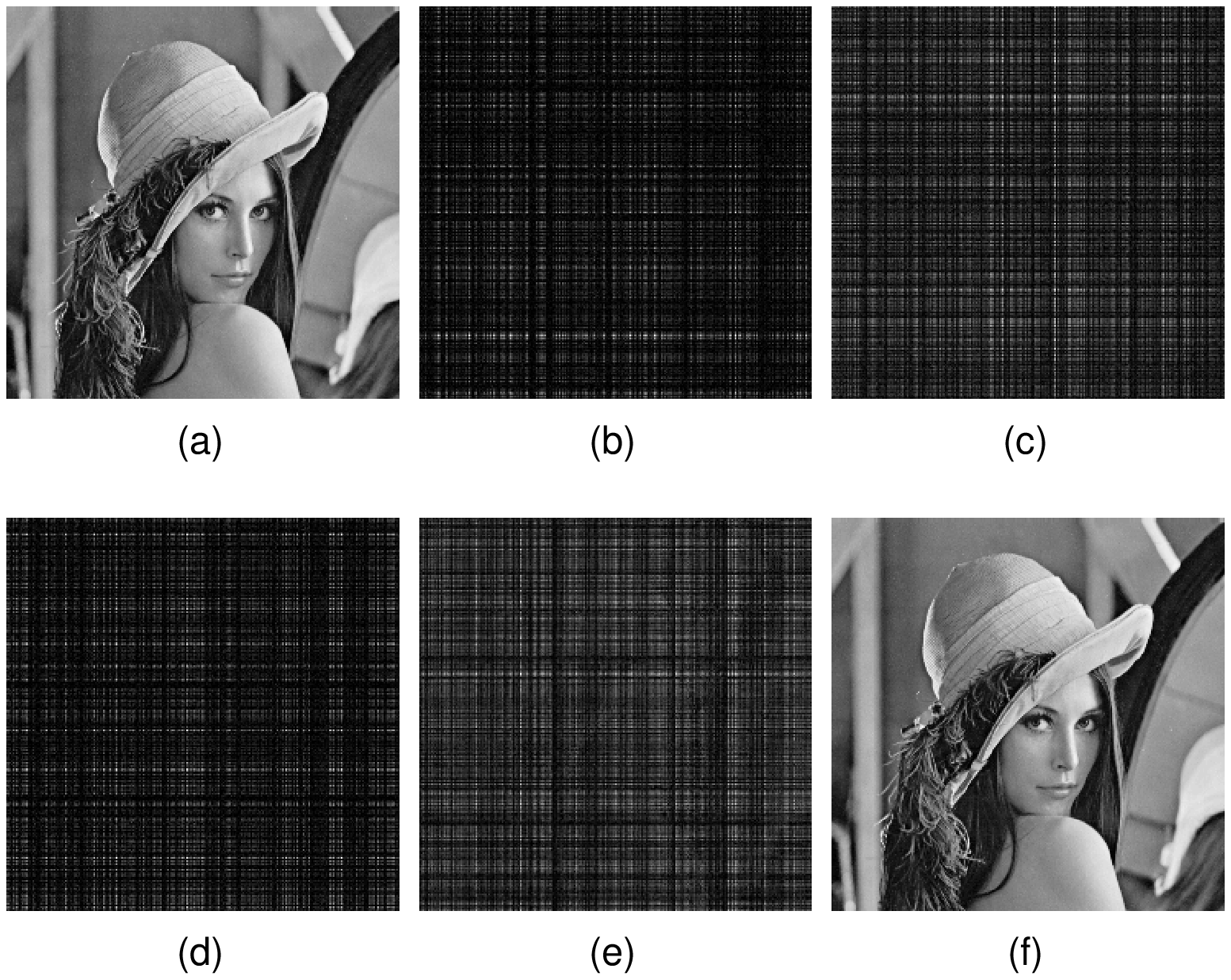}
  \caption{Numerical results of DFRNT with two-dimensional data using a normally
  distributed random matrix. DFRNT serves as an image encryption and
decryption algorithm here. (a) The original image, (b) encrypted
image with $\alpha=0.5$, (c) encrypted image with $\alpha=0.8$,
(d) decryption result for image (b) with a different random
matrix, (e) decryption result for image (b) with $\alpha=-0.502$,
and (f) the correct decryption of the image. }
  \end{figure}

\begin{figure}[h]
\includegraphics[width=16.00cm]{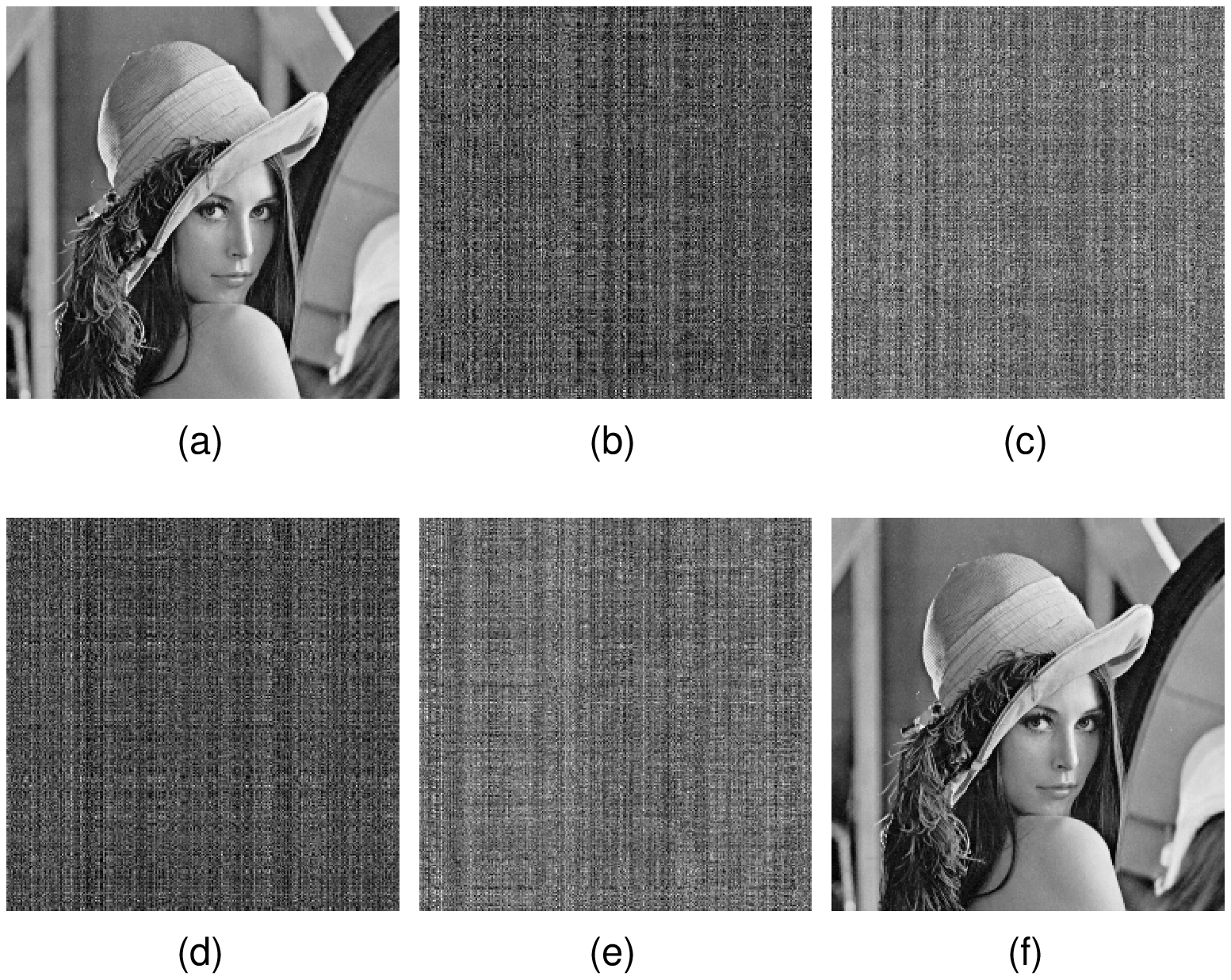}
  \caption{Numerical results of DFRNT with two-dimensional data using a
  uniformly distributed random matrix. DFRNT serves as an image encryption and
decryption algorithm here. (a) The original image, (b) encrypted
image with $\alpha=0.5$, (c) encrypted image with $\alpha=0.8$,
(d) decryption result for image (b) with a different random
matrix, (e) decryption result for image (b) with $\alpha=-0.503$,
and (f) the correct decryption of the image. }
  \end{figure}


\newpage

\begin{thebibliography}{10}

\bibitem{Goodman} J. W. Goodman,  {\it Introduction to Fourier
Optics} (New York: McGraw-Hill), 1968.

\bibitem{Nielsen} M. A. Nielsen and I. Chuang,
{\it Quantum computation and quantum information} (Cambridge:
Cambridge Uni. Press), 2000.

\bibitem{wavelet} C. K. Chui,  {\it An intoduction to wavelets}
(Academic Press, Inc.), 1992.

\bibitem{namias} V. Namias, The fractional Fourier order
Fourier transform and its application to quantum mechanics,
\textit{J. Inst. Maths Appl.} {\bf 25}, (1980) 241.

\bibitem{mcbridge} A. C. McBrdige and F. H. Kerr, On Namias's
fractional Fourier transforms, \textit{IMA J. Appl. Math.} {\bf 39},
(1987) 159.

\bibitem{ozaktas1}D. Mendlovic and H. M. Ozaktas, Fractional
Fourier transfroms and their optical implementation: I, J. Opt. Soc.
Am. {\bf A10}, (1993) 1875.

\bibitem{lohmann}A. W. Lohmann, Image rotation, Wigner
rotation, and the fractional order Fourier transform, \textit{J.
Opt. Soc. Am.} {\bf A10}, (1993) 2181.

\bibitem{ozaktas2}D. Mendlovic, H. M. Ozaktas and A. W. Lohmann,
Graded-index fibers, Wigner-distibution functions, and the
fractional Fourier transform, {\it Appl. Opt.} {\bf 33}, (1994)
6188.

\bibitem{almeida} L. B. Almeida, The fractional Fourier transform
and time-frequency representation, \textit{IEEE Trans. Sig. Proc.}
{\bf 42}, (1994) 3084.

\bibitem{ozaktas3}H. M. Ozaktas, Z. Zalevsky, and M. A. Kutay,
{\it The fractional Fourier transform with applications in optics
and signal processing}, (New York: John Wiley \& Sons), 2000.

\bibitem{dfrft}S. C. Pei and M. H. Yeh, Improved discrete
fractional Fourier transform, \textit{Opt. Lett.} \textbf{22},
(1997) 1407.

\bibitem{dfrct}S. C. Pei and M. H. Yeh, The discrete fractional
cosine and sine transforms, \textit{IEEE Trans. Sig. Proc.}
\textbf{49}, (2001) 1198.

\bibitem{encryp1}S. Liu, L. Yu, and B. Zhu, Optical image encryption by cascaded
fractional Fourier transforms with random phase filtering,
\textit{Opt. Commun.} \textbf{187}, (2001) 57.

\bibitem{encryp2}S. Liu , Q. Mi, and B. Zhu. Optical image encryption with multi-stage
and multi-channel fractional Fourier domain filtering, \textit{Opt.
Lett.} \textbf{26}, (2001) 1242.

\bibitem{genfrft}S. Liu, J. Jiang, Y. Zhang and J. Zhang,
Generalized fractional Fourier transforms, \textit{J. Phys. A: Math
\& Gen.} \textbf{30}, (1997) 973.

\bibitem{dft} B. W. Dickinson  and K. Steiglitz, Eigenvectors and functions of the
discrete Fourier transform \textit{IEEE Trans. Acoust. Speech. \&
Sig. Proc.} \textbf{ASSP-30}, (1982) 25.

\bibitem{mccellan} J. H. McCellan and T. W. Parks, Eigenvalue and
eigenvector decomposition of the discrete Fourier transform,
\textit{IEEE Trans. Audio Electroacoustics} \textbf{20}, (1972) 66.

\end{thebibliography}
\end{document}